\documentclass{article}

\RequirePackage[top=1.5in,bottom=1.5in,left=1.5in,right=1.5in]{geometry}
\RequirePackage{subcaption}
\RequirePackage[labelfont=bf]{caption}

\RequirePackage{amsmath,amsfonts,amssymb,amsthm}
\RequirePackage{graphicx}
\RequirePackage[colorlinks=true]{hyperref}
\RequirePackage{appendix}
\RequirePackage{url}
\usepackage[
    backend=biber,
    url=true,
    doi=true,
    eprint=true
]{biblatex}
\newcommand{\T}{\mathsf{T}}
\newcommand{\Reals}{\ensuremath{\mathbb{R}}}
\newcommand{\Naturals}{\ensuremath{\mathbb{N}}}

\newcommand{\Dir}{\text{Dir}}
\newcommand{\Mult}{\text{Mult}}
\newcommand{\TMult}{\text{TruncMult}}

\begin{document}
    \begin{center}
        \Large
        Dirichlet Posterior Sampling\\
        with Truncated Multinomial Likelihoods\\
        \vspace{0.2in}
        \normalsize
        Matthew James Johnson, MIT\\
        Alan S. Willsky, MIT\\
        \vspace{0.2in}
        \footnotesize
        Revised \today
    \end{center}

    \section{Overview}
    This document considers the problem of drawing samples from posterior
    distributions formed under a Dirichlet prior and a truncated multinomial
    likelihood, by which we mean a Multinomial likelihood function where we
    condition on one or more counts being zero a priori.
    An example is the distribution with density
    \begin{align}
        p(\pi | m, \alpha)
        &\propto \underbrace{\prod_i \pi_i^{\alpha_i - 1}}_{\text{prior}}
        \cdot \underbrace{\left(\prod_{i \neq 1} \left( \frac{\pi_i}{1 - \pi_1} \right)^{m_{1i}}
            \right)\left(\prod_{i \neq 2} \left( \frac{\pi_i}{1 - \pi_2} \right)^{m_{2i}}
            \right)}_{\text{likelihood}}
    \end{align}
    where $\pi \in \Delta := \left\{ x \in \Reals_+^n : \sum_i x_i = 1
    \right\}$, $\alpha \in \Reals_+^n$, and $\Reals_+ = \{ x \in \Reals : x
        \geq 0 \}$. We say the likelihood function has two \emph{truncated}
    terms because each term corresponds to a multinomial likelihood defined on
    the full parameter $\pi$ but conditioned on the event that observations
    with a certain label are removed from the data.

    Sampling this posterior distribution is of interest in inference algorithms
    for hierarchical Bayesian models based on the Dirichlet distribution or the
    Dirichlet Process, particularly the sampling algorithm for the Hierarchical
    Dirichlet Process Hidden Semi-Markov Model (HDP-HSMM)
    \cite{johnson2011hdphsmm} which must draw samples from such a distribution.

    We provide an auxiliary variable (or data augmentation) \cite{van2001art}
    sampling algorithm that is easy to implement, fast both to mix and to
    execute, and easily scalable to high dimensions. This document will
    explicitly work with the finite Dirichlet distribution, but the sampler
    immediately generalizes to the Dirichlet Process case based on the
    Dirichlet Process's definition in terms of the finite Dirichlet
    distribution and the Komolgorov extension theorem
    \cite{orbanz2010construction}.

    Section~\ref{sec:problem} explains the problem in greater detail.
    Section~\ref{sec:auxsampler} provides a derivation of our sampling
    algorithm.  Finally, Section~\ref{sec:experiments} provides numerical
    experiments in which we demonstrate the algorithm's significant advantages
    over a generic Metropolis-Hastings sampling algorithm.

    Sampler code and functions to generate each plot in this document are
    available at
    \url{https://github.com/mattjj/dirichlet-truncated-multinomial}.

    \section{Problem Description}
    \label{sec:problem}
    We say a vector $\pi \in \Delta$ is Dirichlet-distributed with
    parameter vector $\alpha \in \Reals_+^n$ if it has a density
    \begin{align}
        p(\pi | \alpha) &= \frac{\Gamma \left( \sum_{i=1}^n \alpha_i
            \right)}{\prod_{i=1}^n \Gamma(\alpha_i)} \prod_{i=1}^n
        \pi_i^{\alpha_i - 1}\\
        &=: \Dir(\pi | \alpha )
    \end{align}
    with respect to Lebesgue measure. The Dirichlet distribution and its
    generalization to arbitrary probability spaces, the Dirichlet Process, are
    common in Bayesian statistics and machine learning models. It is most often
    used as a prior over finite probability mass functions, such as the faces
    of a die, and paired with the multinomial likelihood, to which it is
    conjugate, viz.
    \begin{align}
        \Dir(\pi | \alpha) \cdot \Mult(m | \pi) &\propto \prod_i
        \pi_i^{\alpha-1} \cdot \prod_i \pi_i^{m_i}\\
        &\propto \prod_i \pi_i^{\alpha_i + m_i - 1}\\
        &\propto \Dir(\pi | \alpha + m).
    \end{align}
    That is, given a count vector $m \in \Naturals_+^n$, the posterior
    distribution is also Dirichlet with an updated parameter vector and,
    therefore, it is easy to draw samples from the posterior.

    However, we consider a modified likelihood function which does not maintain
    the convenient conjugacy property: the \emph{truncated} multinomial
    likelihood, which corresponds to deleting a particular set of counts from
    the count vector $m$ or, equivalently, conditioning on the event that they
    are not observed. The truncated multinomial likelihood where the first
    component is truncated can be written
    \begin{align}
        \TMult_{\{1\}}( m | \pi ) &:= \prod_{i \neq 1} \left( \frac{\pi_i}{1 - \pi_1}
        \right)^{m_i}\\
        &= \Mult( m | \pi, \{ m_1 = 0 \}).
    \end{align}
    In general, any subset of indices  may be truncated; if a set $\mathcal{I}
    \subseteq \{1,2,\ldots,n\}$ is truncated, then we write
    \begin{align}
        \TMult_{\mathcal{I}}(m | \pi) &:= \left( \frac{1}{1-\sum_{i \in \mathcal{I}}
                \pi_i} \right)^{m_{\cdot}} \prod_{i \not\in \mathcal{I}} \pi_i^{m_i}
    \end{align}
    where $m_{\cdot} = \sum_i m_i$. This distribution can arise in hierarchical
    Bayesian models such as the HDP-HSMM \cite{johnson2011hdphsmm}.

\begin{figure}[tp]
        \centering
        \begin{subfigure}{2.1in}
            \includegraphics[width=2in]{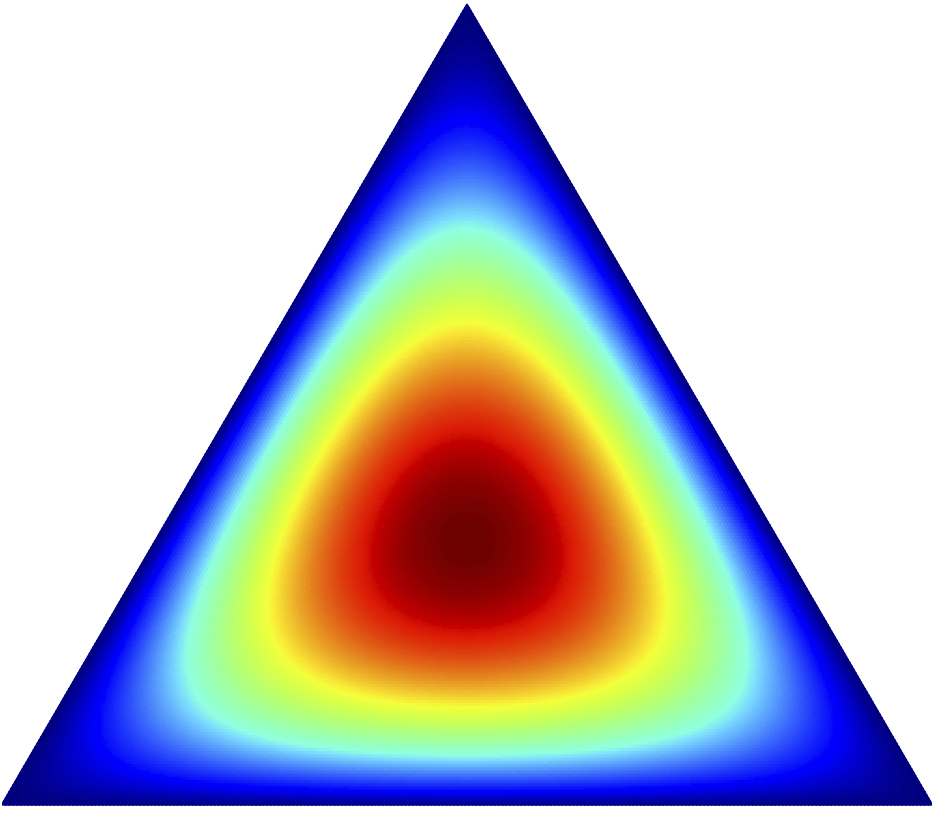}
            \caption{The prior, $\Dir(\pi | \alpha)$.}
        \end{subfigure}\\
        \begin{subfigure}{2.2in}
            \includegraphics[width=2in]{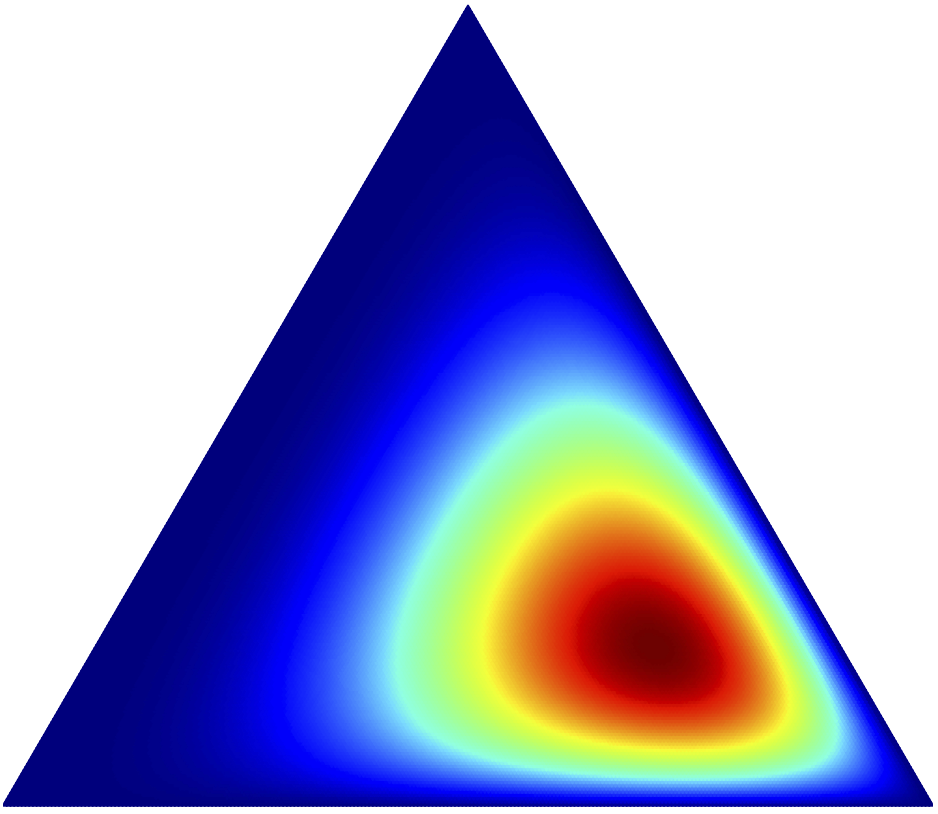}
            \caption{The posterior proportional to $\Dir(\pi | \alpha) \cdot
                \Mult( m | \pi)$.}
        \end{subfigure}\qquad
        \begin{subfigure}{2.2in}
            \includegraphics[width=2in]{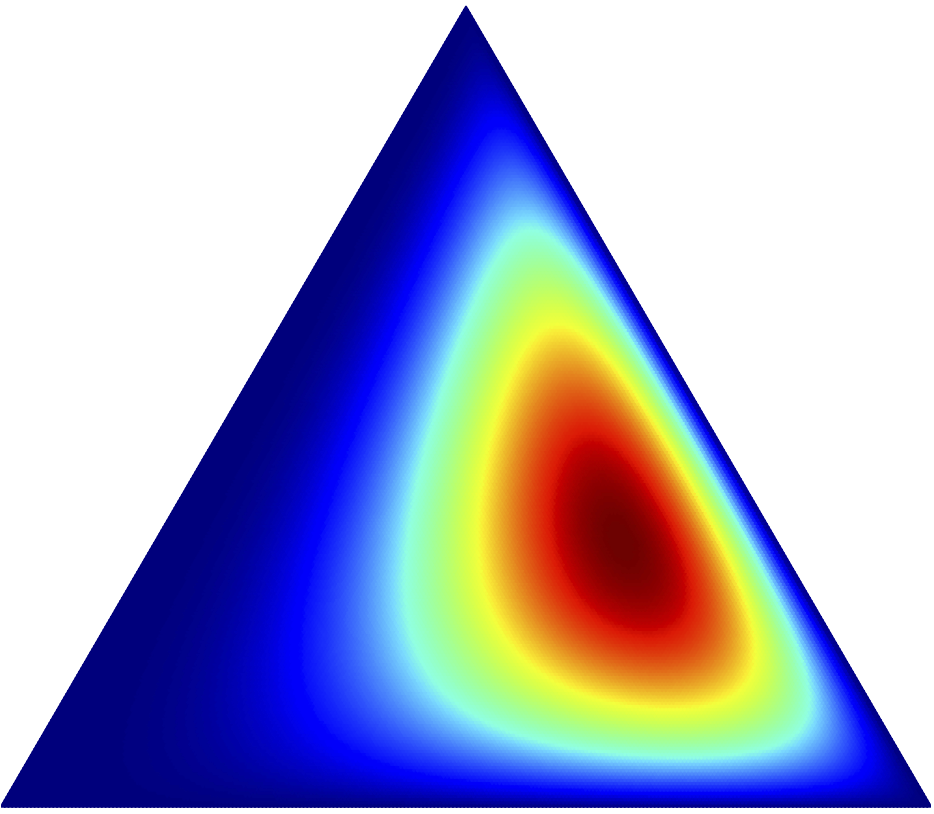}
            \caption{The posterior proportional to
                $\Dir(\pi | \alpha) \cdot \TMult_{\{1\}}( m | \pi)$.}
        \end{subfigure}
        \caption{Projected visualizations of the prior distribution $\Dir(\pi |
            \alpha)$ for $n=3$ and $\alpha = (2,2,2)$ and the associated
            posterior distributions when paired with $\Mult(m | \pi)$ and
            $\TMult_{\{1\}} (m|\pi)$ where $m=(0,2,0)$. In low dimensions, the
            posteriors can be computed via direct numerical integration over a
            discretized mesh.}
        \label{fig:densities_2D}
    \end{figure}

    In the case where the posterior is proportional to a Dirichlet prior and a
    single truncated multinomial likelihood term, the posterior is still simple
    to write down and sample. In this case, we may split the Dirichlet prior
    over $\mathcal{I}$ and its complement $\overline{\mathcal{I}} :=
    \{1,2,\ldots,n\} \setminus \mathcal{I}$; the factor over
    $\overline{\mathcal{I}}$ is conjugate to the likelihood, and so the
    posterior can be written
    \begin{align}
        \Dir(\pi|\alpha) \TMult_{\mathcal{I}}(m|\pi) &\propto
        \Dir \left. \left( \frac{\pi_{\mathcal{I}}}{1 - \sum_{i \in \mathcal{I}} \pi_i} \right| \alpha_{\mathcal{I}} \right)
        \Dir \left. \left( \frac{\pi_{\overline{\mathcal{I}}}}{1 - \sum_{i \in \mathcal{I}} \pi_i} \right| \alpha_{\overline{\mathcal{I}}} +
        m_{\overline{\mathcal{I}}} \right)
    \end{align}
    from which we can easily sample. However, given two or more truncated
    likelihood terms with different truncation patterns, no simple conjugacy
    property holds, and so it is no longer straightforward to construct samples
    from the posterior.  For a visual comparison in the $n=3$ case, see
    Figure~\ref{fig:densities_2D}.

    For the remainder of this document, we deal with the case where there are
    two likelihood terms, each with one component truncated.  The
    generalization of the equations and algorithms to the case where any set of
    components is truncated is immediate.

    \section{An Auxiliary Variable Sampler}
    \label{sec:auxsampler}
    Data augmentation methods are auxiliary variable methods that often provide
    excellent sampling algorithms because they are easy to implement and the
    component steps are simply conjugate Gibbs sampling steps, resulting in
    fast mixing. For an overview, see the survey \cite{van2001art}.

    We can derive an auxiliary variable sampler for our problem by augmenting
    the distribution with geometric random variables $k = (k_1,k_2) =
    (\{k_{1j}\}, \{k_{2j}\})$. That is, we define for $k_{ij}=\{0,1,2,\ldots\}$ a
    new distribution $q$ such that
    \begin{align}
        q(\pi, m, k | \alpha ) &\propto
        \left( \prod_i \pi_i^{\alpha-1} \right)
        \left( \prod_{i \neq 1} \pi_i^{m_{1i}} \right)
        \left( \prod_{i \neq 2} \pi_i^{m_{2i}} \right)
        \left( \prod_{j=1}^{m_{1\cdot}} \pi_1^{k_{1j}} \right)
        \left( \prod_{j=1}^{m_{2\cdot}} \pi_2^{k_{2j}} \right)
            \label{eqn:augmenteddistn}
    \end{align}
    where $\{m_{1i}\}$ and $\{m_{2i}\}$ are sample counts corresponding to each
    likelihood, respectively, and $m_{i\cdot} := \sum_j m_{ij}$. Note that if
    we sum over all the auxiliary variables $k$, we have
    \begin{align}
        \sum_{k} q(\pi,m,k | \alpha) &\propto
        \left( \prod_i \pi_i^{\alpha-1} \right)
        \left( \prod_{i \neq 1} \pi_i^{m_{1i}} \right)
        \left( \prod_{i \neq 2} \pi_i^{m_{2i}} \right)
        \left( \prod_j \sum_{k_{1j}} \pi_1^{k_{1j}} \right)
        \left( \prod_j \sum_{k_{2j}} \pi_2^{k_{2j}} \right)\\
        &= \prod_i \pi_i^{\alpha-1}
        \left( \prod_{i \neq 1} \left( \frac{\pi_i}{1 - \pi_1} \right)^{m_{1i}} \right)
        \left( \prod_{i \neq 2} \left( \frac{\pi_i}{1 - \pi_2} \right)^{m_{2i}} \right) \\
        &\propto p(\pi,m | \alpha)
    \end{align}
    and so if we can construct samples of $\pi,k | m, \alpha$ from the
    distribution $q$ then we can form samples of $\pi|m,\alpha$ according to
    $p$ by simply ignoring the values sampled for $k$.

    We construct samples of $\pi,k | m,\alpha$ by iterating Gibbs steps between
    $k | \pi, m, \alpha$ and $\pi|k,m,\alpha$. We see from
    \eqref{eqn:augmenteddistn} that each $k_{ij}$ in $k | \pi, m, \alpha = k |
    \pi, m$ is independent and distributed according to
    \begin{align}
        q(k_{ij} | \pi, m) &= (1-\pi_i) \pi_i^{k_{ij}}.
    \end{align}
    Therefore, each $k_{ij}$ follows a geometric distribution with success
    parameter $(1-\pi_i)$.

\begin{figure}[tp]
        \centering
        \begin{subfigure}{2in}
            \centering
            \includegraphics[scale=0.5]{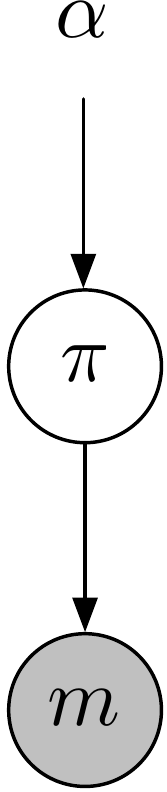}
            \caption{Un-augmented distribution.}
        \end{subfigure}\qquad
        \begin{subfigure}{2in}
            \centering
            \includegraphics[scale=0.5]{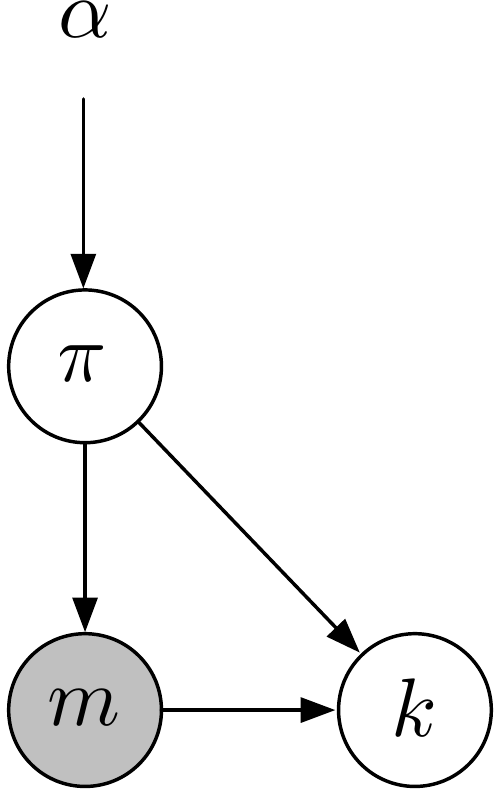}
            \caption{Augmented distribution.}
        \end{subfigure}
        \caption{Graphical models for the un-augmented and augmented
            probability models.}
        \label{fig:augmentation_graphicalmodel}
    \end{figure}

    The distribution of $\pi | k,m,\alpha$ in $q$ is also simple:
    \begin{align}
        q(\pi | m, k, \alpha) &\propto
        \left( \prod_{i} \pi_i^{\alpha_i-1} \right)
        \left( \prod_{i \neq 1} \left(\frac{\pi_i}{1-\pi_1}\right)^{m_{1i}} \right)
        \left( \prod_{i \neq 2} \left(\frac{\pi_i}{1-\pi_2}\right)^{m_{2i}} \right)\\
        &\qquad \cdot \left( \prod_{j=1}^{m_{1\cdot}} (1-\pi_1) \pi_1^{k_{1j}} \right)
        \left( \prod_{j=1}^{m_{2\cdot}} (1-\pi_2) \pi_2^{k_{2j}} \right) \\
        &\propto \Dir(\pi | \alpha + \tilde{m})
    \end{align}
    where $\tilde{m}$ is a set of \emph{augmented} counts including the values
    of $k$. In other words, the Dirichlet prior is conjugate to the augmented
    model. Therefore we can cycle through Gibbs steps in the augmented
    distribution and hence easily produce samples from the desired posterior.
    For a graphical model of the augmentation, see
    Figure~\ref{fig:augmentation_graphicalmodel}.

    \section{Numerical Experiments}
    \label{sec:experiments}
    In this section we perform several numerical experiments to demonstrate the
    advantages provided by the auxiliary variable sampler. We compare to a
    generic Metropolis-Hastings sampling algorithm. For all experiments, when a
    statistic is computed in terms of a sampler chain's samples up to sample
    index $t$, we discard the first $\lfloor \frac{t}{2} \rfloor$ samples and
    use the remaining samples to compute the statistic.

    \paragraph{Metropolis-Hastings Sampler}
    We construct an MH sampling algorithm by using the proposal distribution
    which proposes a new position $\pi'$ given the current position $\pi$ via
    the proposal distribution
    \begin{align}
        p(\pi' | \pi; \beta) &= \Dir(\pi' | \beta \cdot \pi)
    \end{align}
    where $\beta > 0$ is a tuning parameter. This proposal distribution has
    several valuable properties:
    \begin{enumerate}
        \item the mean and mode of the proposals are both $\pi$;
        \item the parameter $\beta$ directly controls the concentration of the
            proposals, so that larger $\beta$ correspond to more local proposal
            samples;
        \item the proposals are naturally confined to the support of the target
            distribution, while alternatives such as local Gaussian proposals
            would not satisfy the MH requirement that the normalization
            constant of the proposal kernel be constant for all starting
            points.
    \end{enumerate}

    In our comparison experiments, we tune $\beta$ so that the acceptance
    probability is approximately $0.24$. 

\begin{figure}
        \centering
        \begin{subfigure}{4in}
            \centering
            \includegraphics[width=4in]{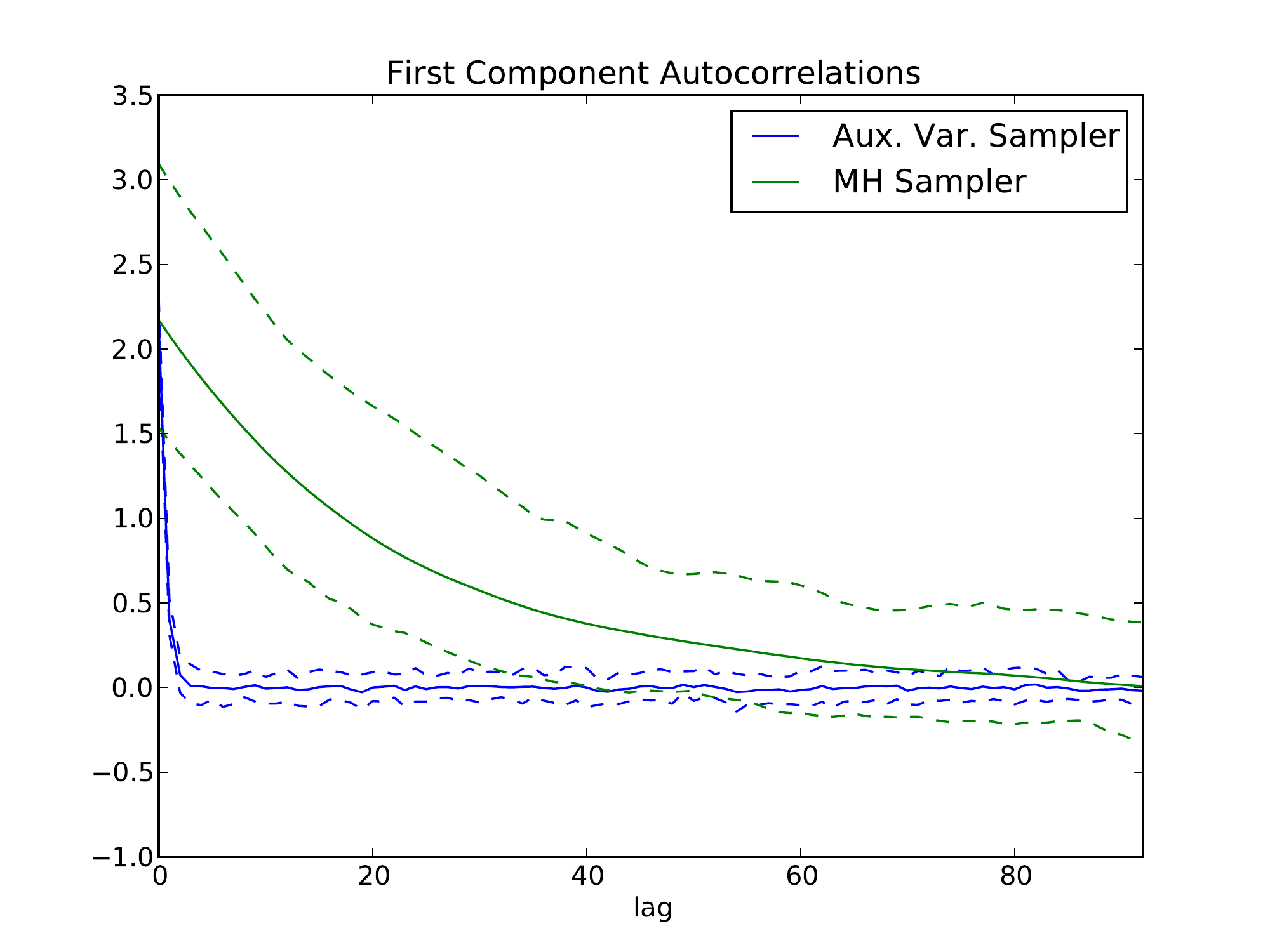}
            \caption{Autocorrelations in the first (truncated) component.}
        \end{subfigure}
        \begin{subfigure}{4in}
            \centering
            \includegraphics[width=4in]{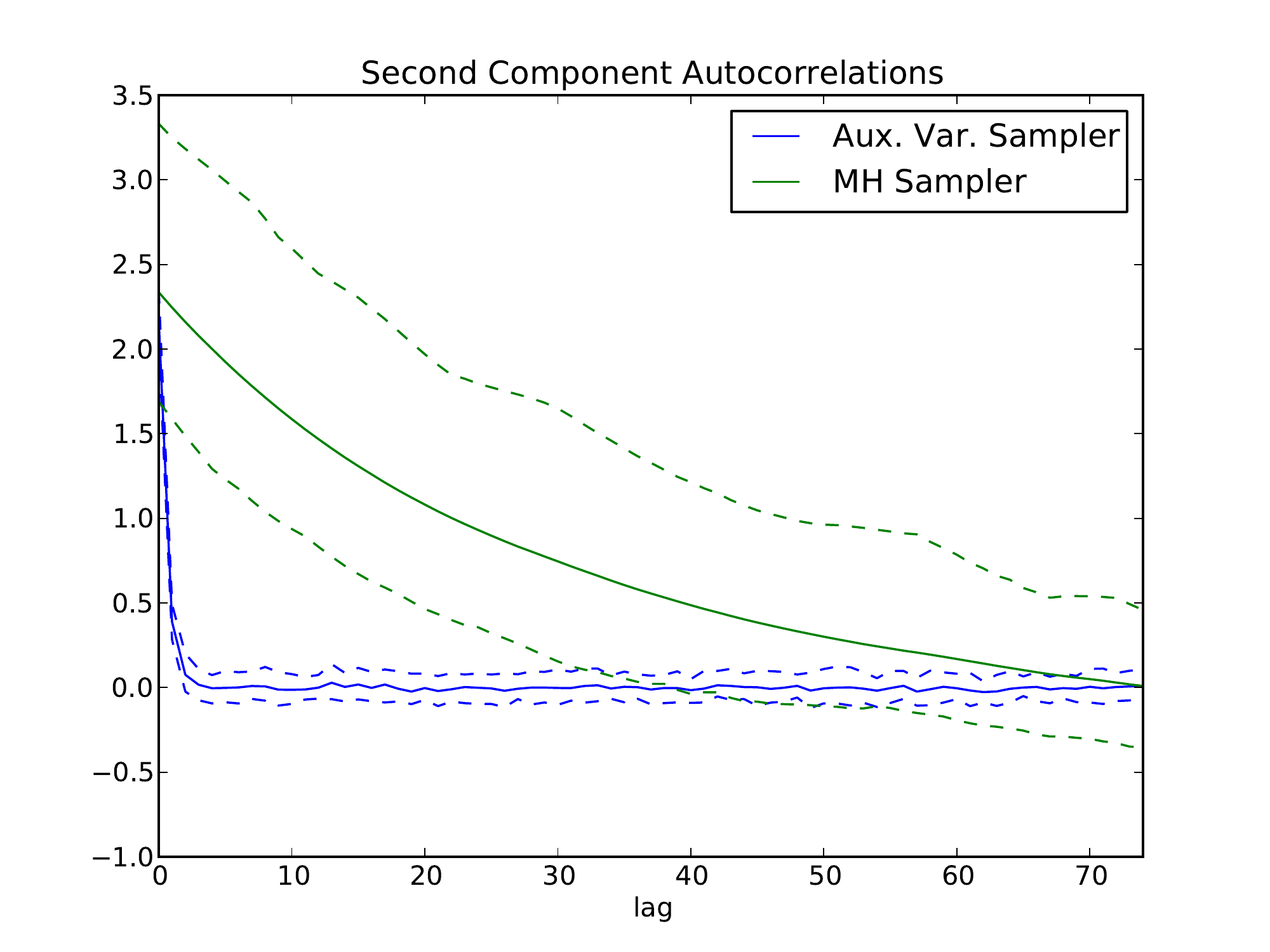}
            \caption{Autocorrelations in the second component.}
        \end{subfigure}
        \caption{Autocorrelations for the auxiliary variable sampler and MH
            sampler chains with $\alpha_i = 2$, $n=10$, $\beta=160$. The solid
            lines show the mean autocorrelation over 50 randomly-initialized
            runs for each sampler, and the dashed lines show the 10th and 90th
            percentile autocorrelation chains over those runs. These plots can
            be reproduced with the function \texttt{autocorrelation} in
            \texttt{figures.py}.}
        \label{fig:autocorrelation}
    \end{figure}

    \paragraph{Sample Chain Autocorrelation}
    In Figure~\ref{fig:autocorrelation} we compare the sample autocorrelation
    of the auxiliary variable sampler and the alternative MH sampler for
    several lags with $n=10$. The reduced autocorrelation that is typical in
    the auxiliary variable sampler chain is indicative of faster mixing.

    \paragraph{The $\widehat{R}$ Multivariate Potential Scale Reduction Factor}
    The $\widehat{R}$ statistic, also called the Multivariate Potential Scale
    Reduction Factor (MPSRF), was introduced in \cite{brooks1998general} and is
    a natural generalization of the scalar Scale Reduction Factor, introduced
    in \cite{gelman1992inference} and discussed in
    \cite[p.~296]{gelman2004bayesian}. As a function of multiple independent
    sampler chains, the statistic compares the between-chain sample covariance
    matrix to the within-chain sample covariance matrix to measure mixing;
    good mixing is indicated by empirical convergence to the statistic's
    asymptotic value of unity.

    Specifically, loosely following the notation of \cite{brooks1998general},
    with $\psi_{jt}^{(i)}$ for denoting the $i$th element of the parameter
    vector in chain $j$ at time $t$ (with $i=1,\ldots,n$, $j=1,\ldots,M$, and
    $t=1,\ldots,T$), to compute the $n$-dimensional MPSRF we form
    \begin{align}
        \widehat{V} &= \frac{T-1}{T} W + \left( 1 + \frac{1}{M} \right) B/T
    \end{align}
    where
    \begin{align}
        W &= \frac{1}{M(T-1)} \sum_{j=1}^{M} \sum_{t=1}^T (\psi_{jt} -
        \bar{\psi}_{j \cdot}) (\psi_{jt} - \bar{\psi}_{j \cdot})^\T\\
        B/T &= \frac{1}{M-1} \sum_{j=1}^M (\bar{\psi}_{j \cdot} -
        \bar{\psi}_{\cdot \cdot}) (\bar{\psi}_{j \cdot} - \bar{\psi}_{\cdot
            c\cdot})^\T.
    \end{align}
    The MPSRF itself is then defined when $W$ is full-rank as \cite[Eq. 4.1 and Lemma
    1]{brooks1998general}
    \begin{align}
        \widehat{R} &:= \sup_{v \in \Reals^n} \frac{v^\T \widehat{V} v}{v^\T W v} \\
                    &=\lambda_{\text{max}} \left( W^{-1} \widehat{V} \right) \\
                    &= \lambda_{\text{max}} \left( W^{-\frac{1}{2}} \widehat{V}
                        W^{\frac{1}{2}} \right)
    \end{align}
    where $\lambda_{\text{max}}(A)$ denotes the eigenvalue of largest modulus
    of the matrix $A$ and the last line follows because conjugating by
    $W^{\frac{1}{2}}$ is a similarity transformation.  Equivalently (and
    usefully for computation), we must find the largest solution $\lambda$ to
    $\det(\lambda W - \widehat{V})=0$.

    However, as noted in \cite[p.~446]{brooks1998general}, the measure is
    incalculable when $W$ is singular, and because our samples are constrained
    to lie in the simplex in $n$ dimensions, the matrices involved will have
    rank $n-1$. Therefore when computing the $\widehat{R}$ statistic,
    we simply perform the natural Euclidean orthogonal projection to the
    $(n-1)$-dimensional affine subspace on which our samples lie; specifically,
    we define the statistic in terms of $Q^\T \widehat{V} Q$ and $Q^\T W Q$,
    where $Q$ is an $n \times (n-1)$ matrix such that $Q^\T Q = I_{(n-1)}$ and
    \begin{align}
        Q R &=
        \begin{pmatrix}
            -(n-1) & 1      & \cdots & 1 & 1     \\
            1      & -(n-1) & \cdots & 1 & 1     \\
                   &        & \vdots &   &       \\
            1      & 1      & \cdots & 1 & -(n-1)\\
            1      & 1      & \cdots & 1 & 1     \\
        \end{pmatrix}
    \end{align}
    for upper-triangular $R$ of size $(n-1) \times (n-1)$.

\begin{figure}
        \centering
        \begin{subfigure}{4in}
            \centering
            \includegraphics[width=4in]{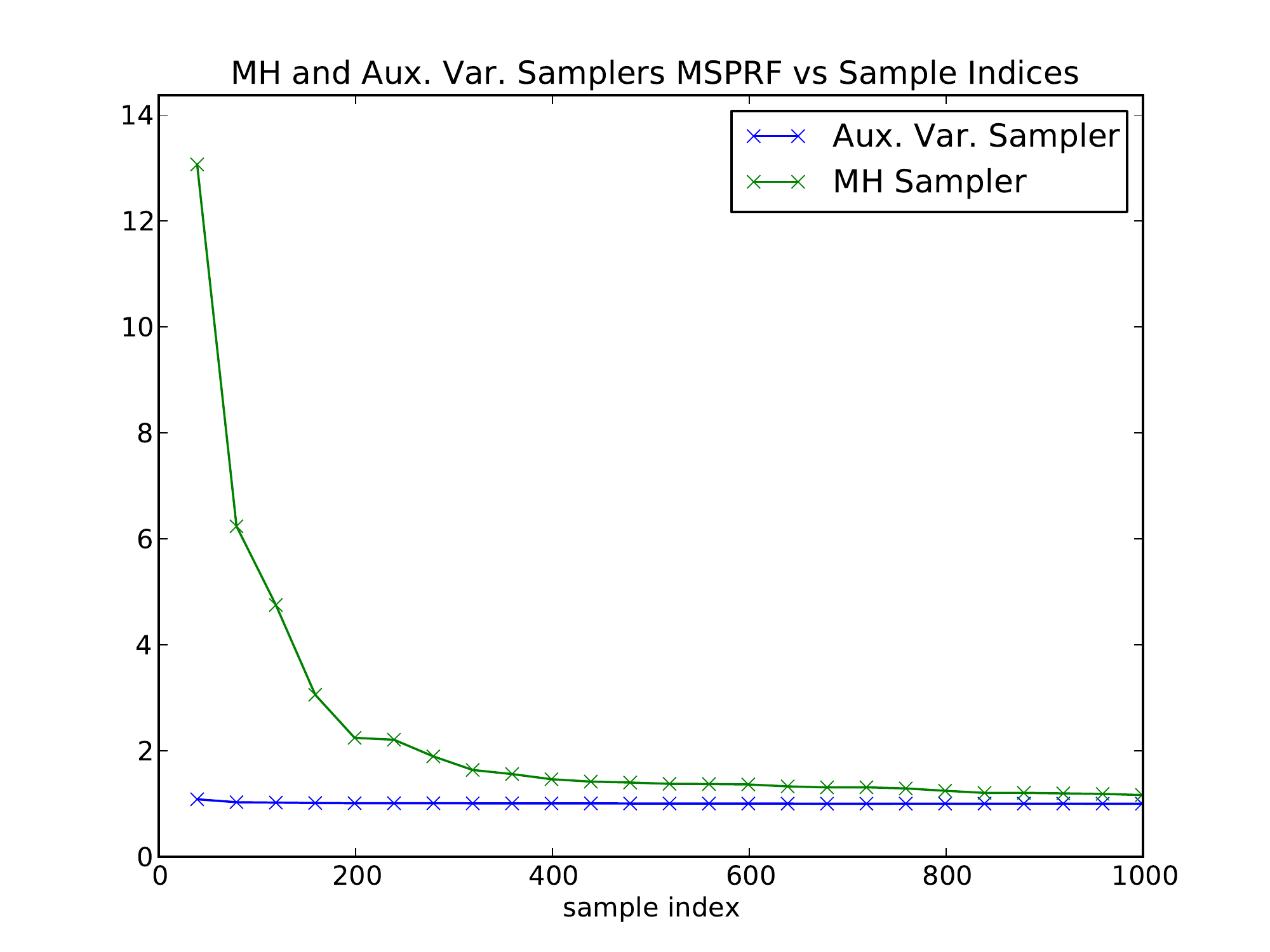}
            \caption{The horizontal axis is the sample index.}
        \end{subfigure}
        \begin{subfigure}{4in}
            \centering
            \includegraphics[width=4in]{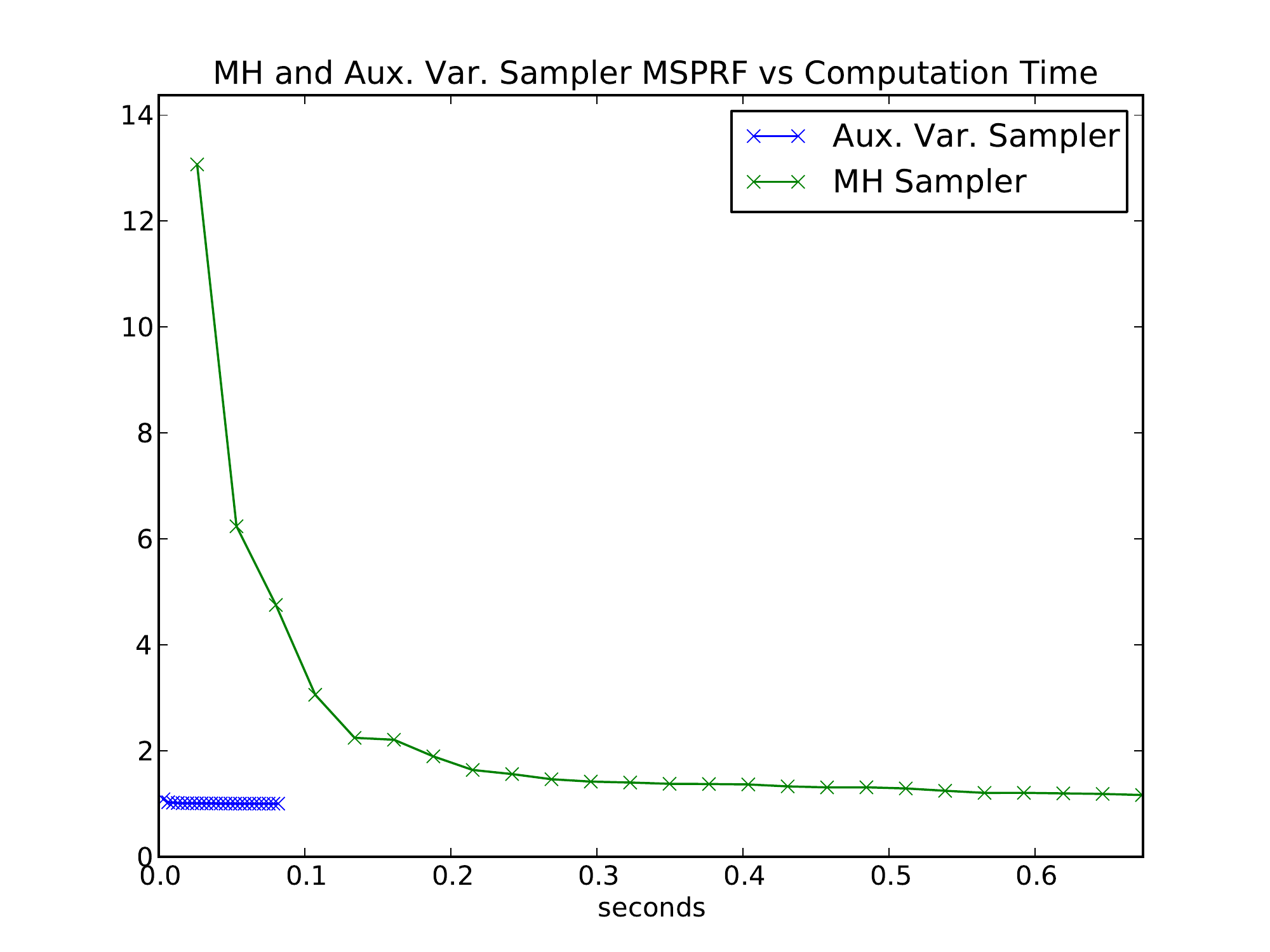}
            \caption{The horizontal axis is elapsed time.}
        \end{subfigure}
        \caption{The $\widehat{R}$ Multivariate Potential Scale Reduction Factor
            \cite{brooks1998general} for the auxiliary variable sampler and MH
            sampler with $\alpha_i = 2$, $n=10$, and $\beta=160$, with
            horizontal axes scaled by sample index and elapsed time. For each
            sampler, 5000 samples were drawn for each of 50
            randomly-initialized runs, and the MPSRF was computed at 25
            equally-spaced intervals. These plots can be reproduced with the
            function \texttt{Rhatp} in \texttt{figures.py}.}
        \label{fig:msprf_10D}
    \end{figure}

\begin{figure}
        \centering
        \begin{subfigure}{4in}
            \centering
            \includegraphics[width=4in]{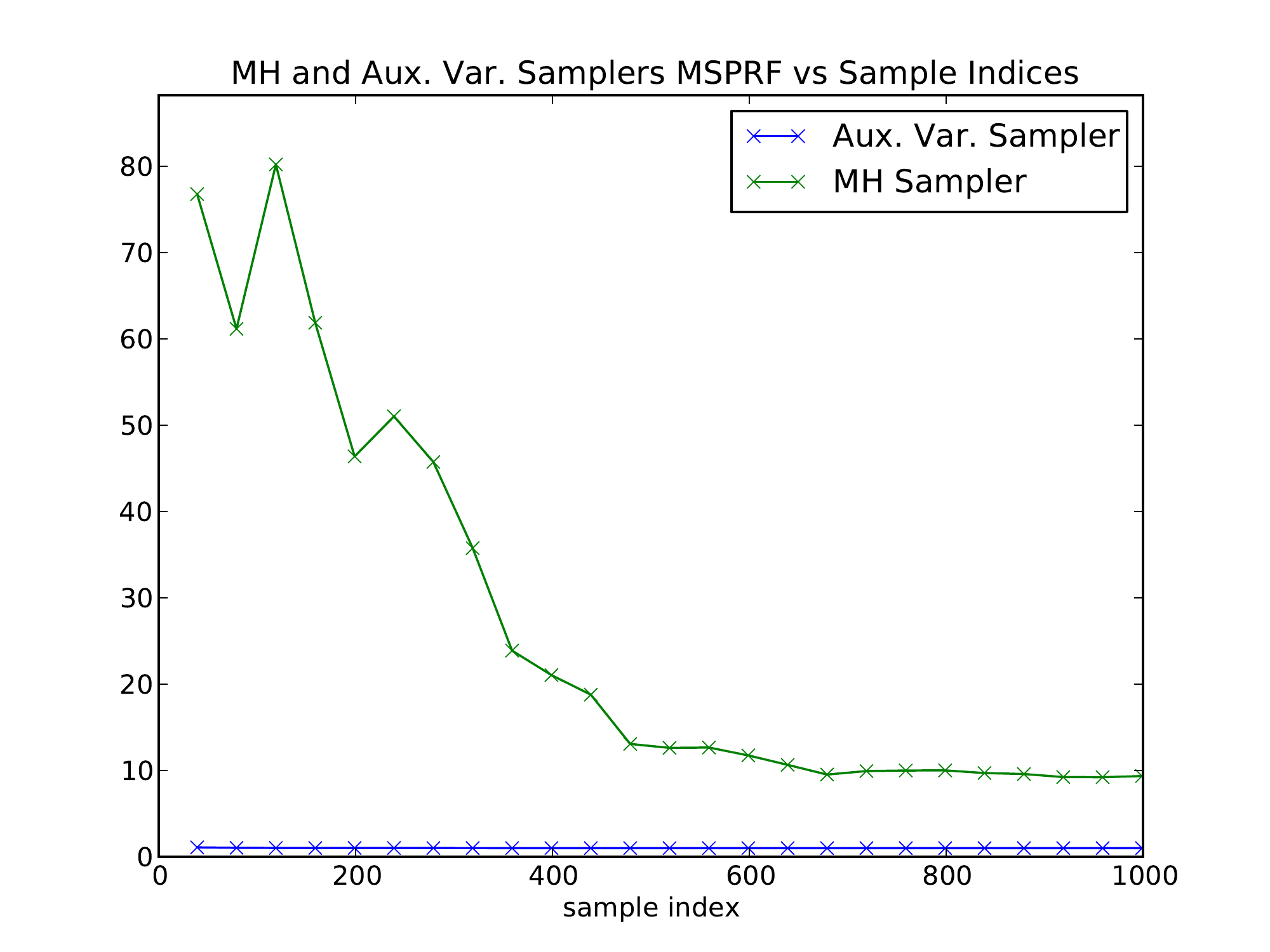}
            \caption{The horizontal axis is the sample index.}
        \end{subfigure}
        \begin{subfigure}{4in}
            \centering
            \includegraphics[width=4in]{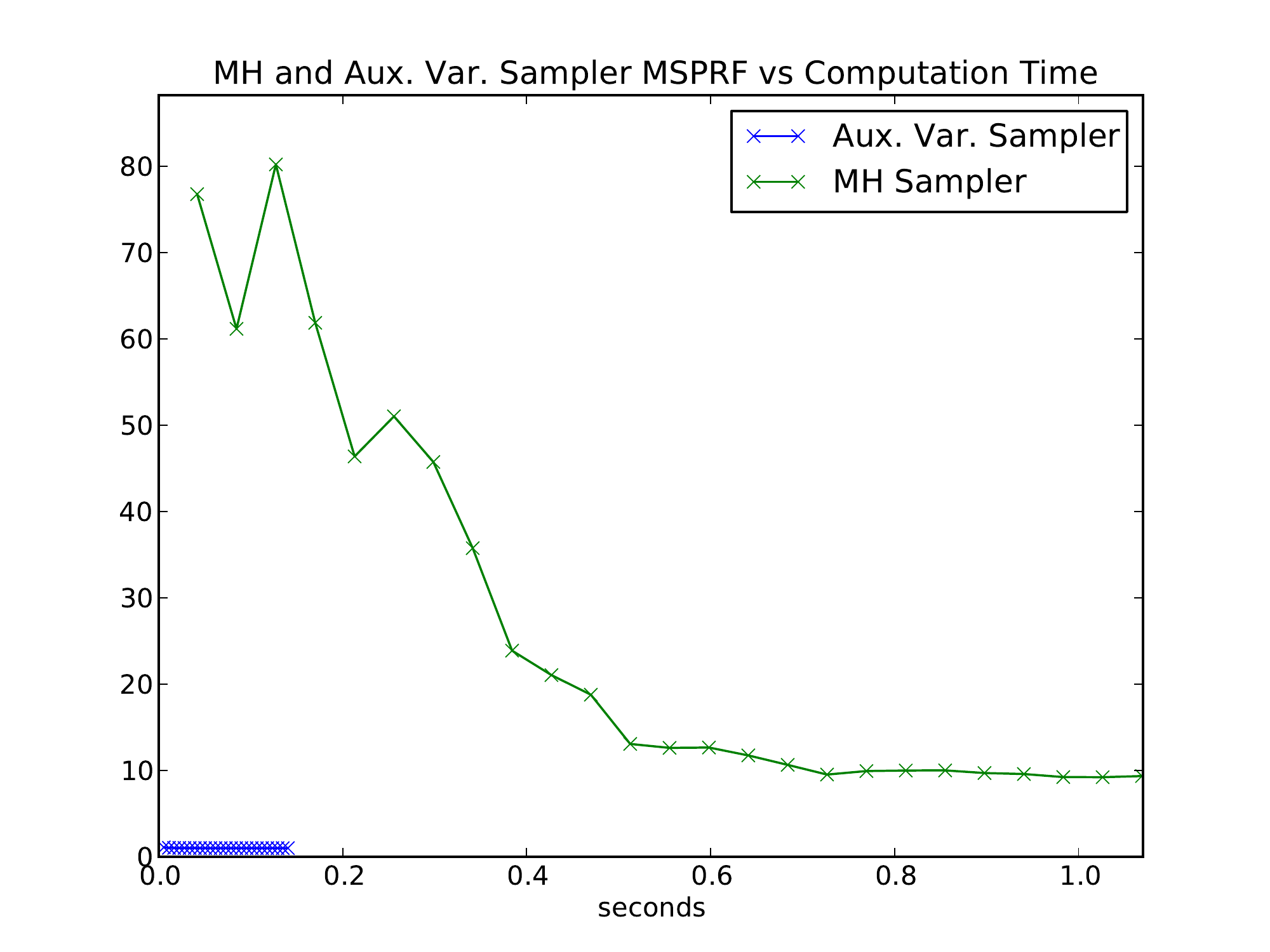}
            \caption{The horizontal axis is elapsed time.}
        \end{subfigure}
        \caption{The $\widehat{R}$ Multivariate Potential Scale Reduction Factor
            \cite{brooks1998general} for the auxiliary variable sampler and MH
            sampler with $\alpha_i = 2$, $n=20$, and $\beta=160$, with
            horizontal axes scaled by sample index and elapsed time. For each
            sampler, 5000 samples were drawn for each of 50
            randomly-initialized runs, and the MPSRF was computed at 25
            equally-spaced intervals. These plots can be reproduced with the
            function \texttt{Rhatp} in \texttt{figures.py}.}
        \label{fig:msprf_20D}
    \end{figure}

    Figure~\ref{fig:msprf_10D} shows the MPSRF of both samplers computed over
    50 sample chains for $n=10$ dimensions, and Figure~\ref{fig:msprf_20D}
    shows the even greater performance advantage of the auxiliary variable
    sampler in higher dimensions.

\begin{figure}
        \centering
        \begin{subfigure}{2.5in}
            \centering
            \includegraphics[width=2.5in]{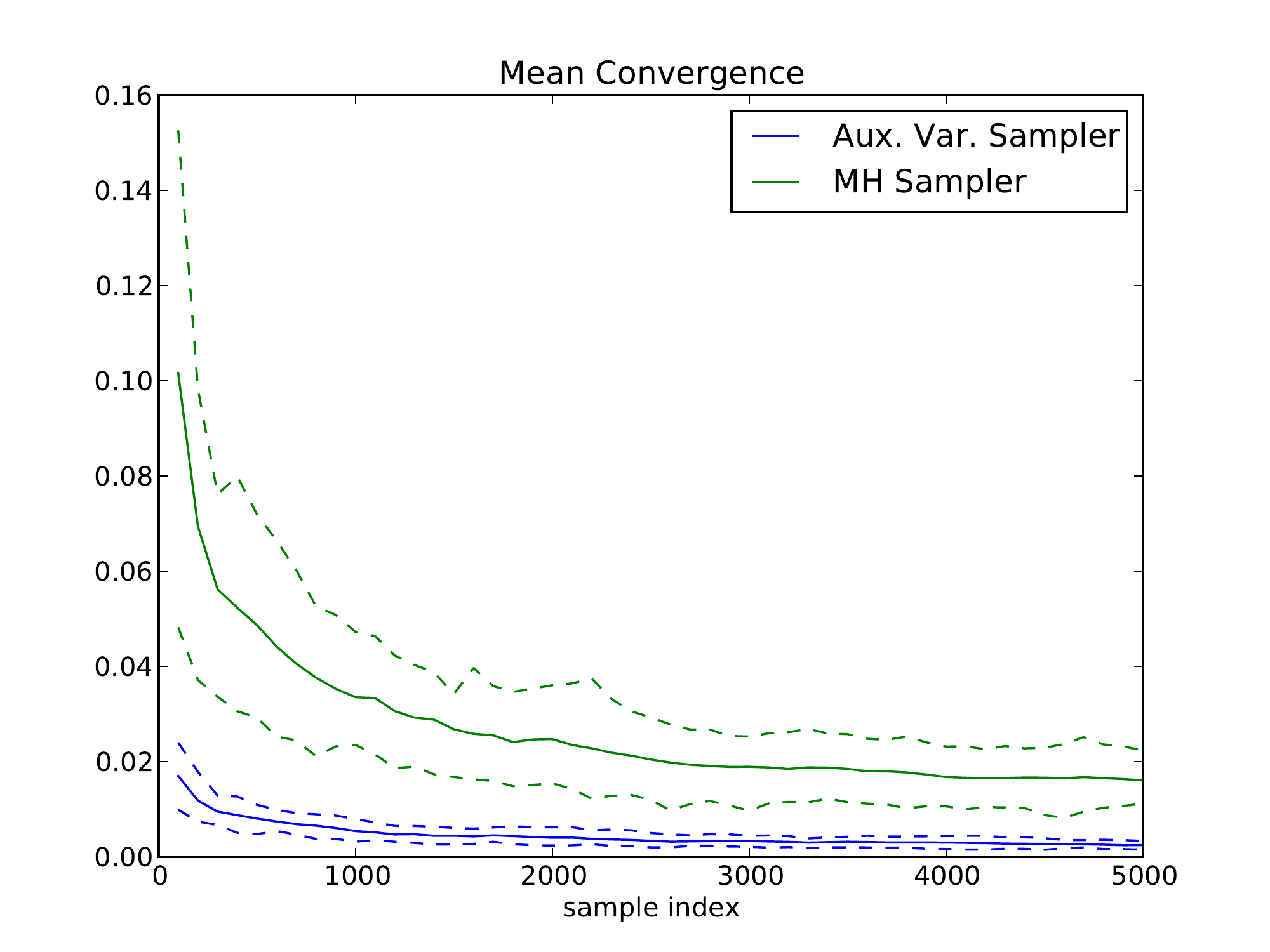}
            \caption{$\ell_2$ error of component-wise mean estimate vs sample index.}
        \end{subfigure}\qquad
        \begin{subfigure}{2.5in}
            \centering
            \includegraphics[width=2.5in]{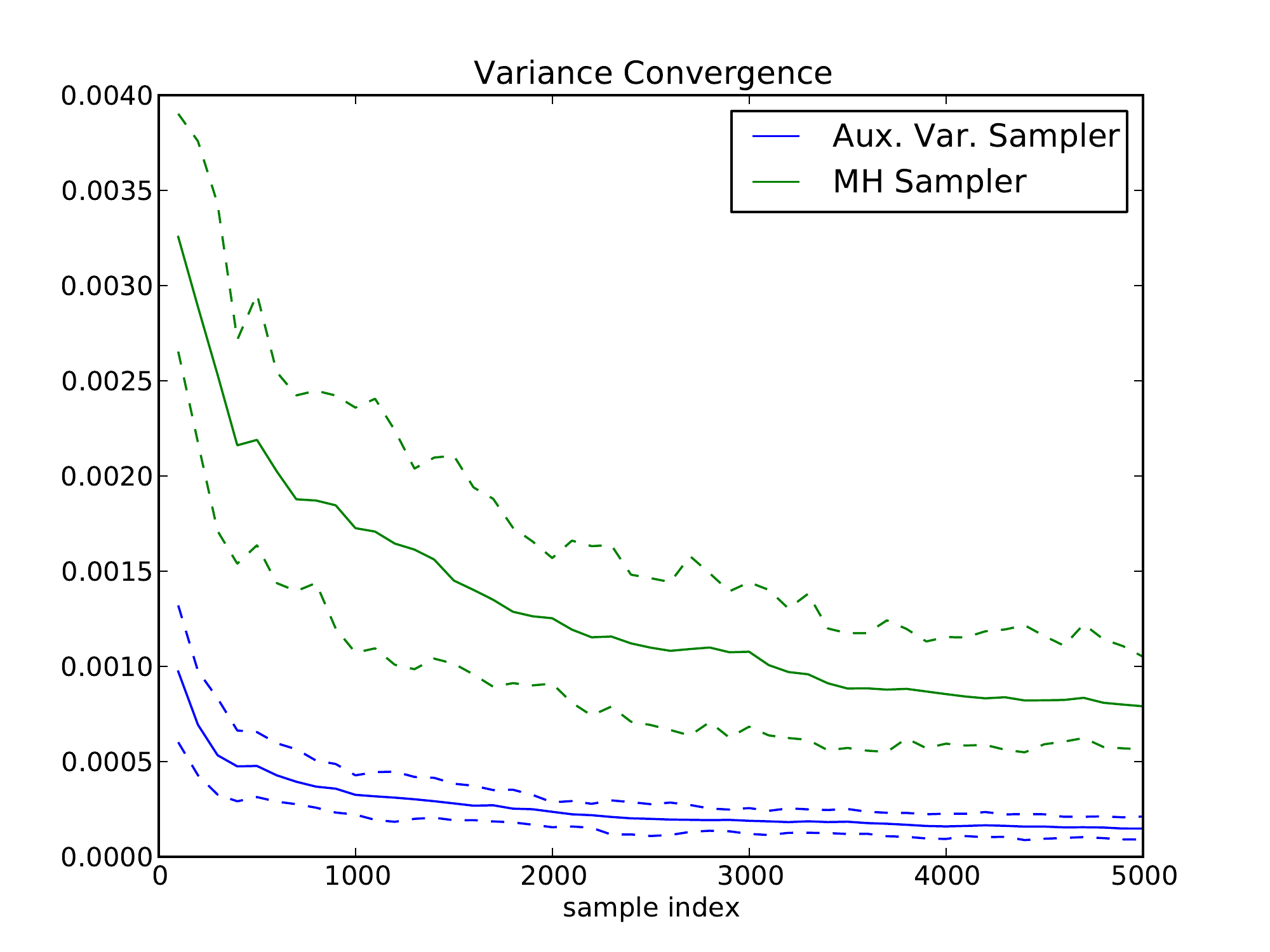}
            \caption{$\ell_2$ error of component-wise variance estimate vs sample index.}
        \end{subfigure}
        \begin{subfigure}{2.5in}
            \centering
            \includegraphics[width=2.5in]{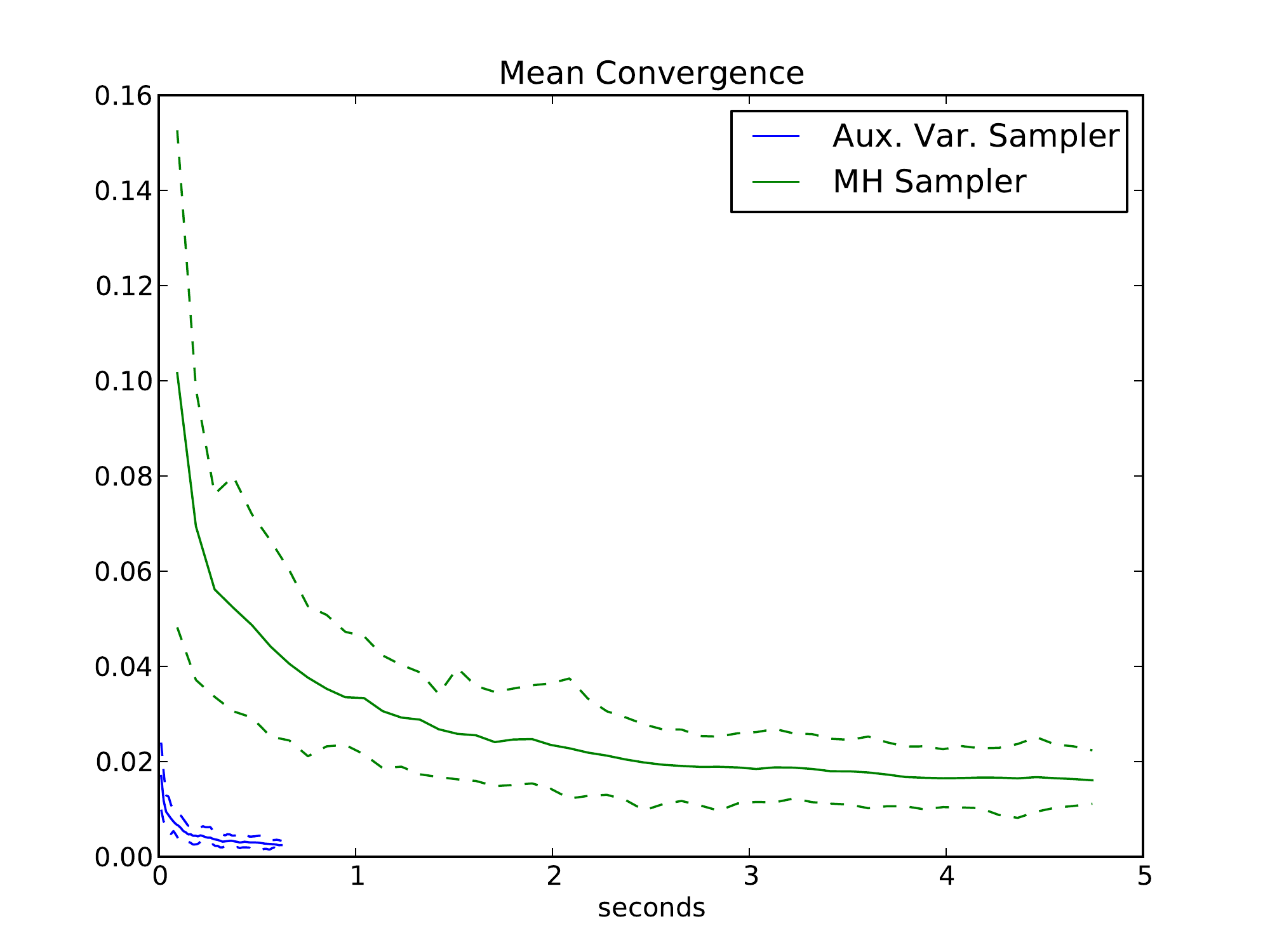}
            \caption{$\ell_2$ error of component-wise mean estimate vs elapsed time.}
        \end{subfigure}\qquad
        \begin{subfigure}{2.5in}
            \centering
            \includegraphics[width=2.5in]{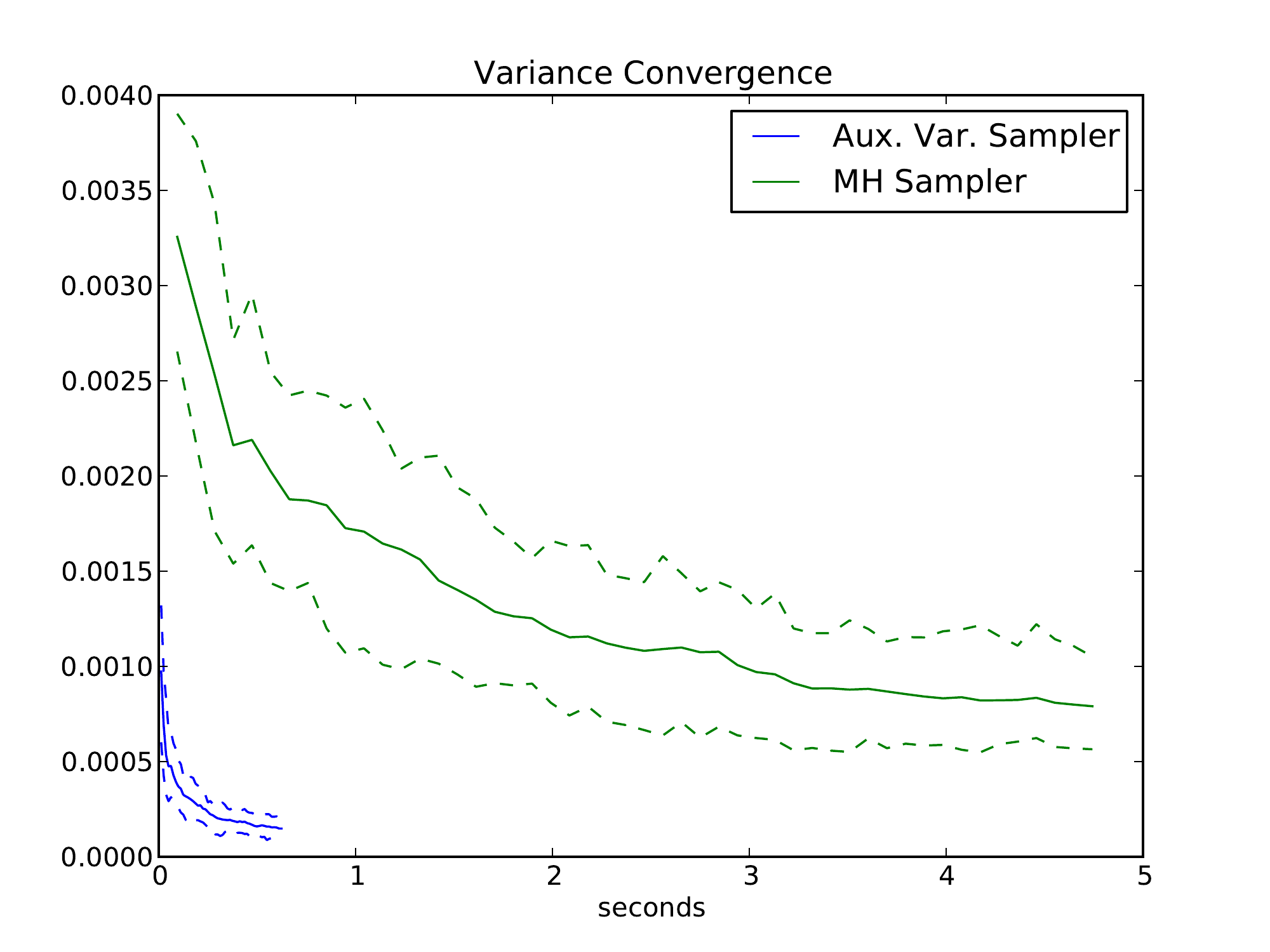}
            \caption{$\ell_2$ error of component-wise variance estimate vs elapsed time.}
        \end{subfigure}
        \caption{Component-wise statistic convergence for the auxiliary
            variable sampler and MH sampler with $\alpha_i=2$, $n=10$, and
            $\beta=160$, with horizontal axes scaled by sample index and
            elapsed time. For each sampler, 5000 samples were drawn for each of
            50 randomly-initialized runs. The $\ell_2$ distances from estimated
               ``true'' parameters are plotted, with the solid lines
               corresponding to the mean error and the dashed lines
               corresponding to 10th and 90th percentile errors. These plots
               can be reproduced with the function
               \texttt{statistic\_convergence} in \texttt{figures.py}.}
        \label{fig:statisticconvergence}
    \end{figure}

    \paragraph{Statistic Convergence}
    Finally, we show the convergence of the component-wise mean and variance
    statistics for the two samplers. We estimated the true statistics by
    forming estimates using samples from 50 independent chains each with 5000
    samples, effectively using 250000 samples to form the estimates. Next, we
    plotted the $\ell_2$ distance between these ``true'' statistic vectors and
    those estimated at several sample indices along the 50 runs for each of the
    sampling algorithms. See the plots in
    Figure~\ref{fig:statisticconvergence}.

    \clearpage
    \printbibliography

\end{document}